\documentclass[preprint,12pt]{elsarticle}

\usepackage{amssymb}
\usepackage{hyperref}
\usepackage{booktabs}
\usepackage{float}
\usepackage{caption}

\journal{Nuclear Physics B}
\begin{document}

\begin{frontmatter}



\title{Pubic Symphysis-Fetal Head Segmentation Using Pure Transformer with Bi-level Routing Attention}


\author{Pengzhou Cai}
\author{Lu Jiang}
\author{Yanxin Li}
\author{Libin Lan\corref{cor1}}
\cortext[cor1]{Corresponding author}
\ead{lanlbn@cqut.edu.cn}
\address{College of Computer Science and Engineering
            ,Chongqing University of Technology, 
            Chongqing,
            400054, 
            China}

\begin{abstract}
In this paper, we propose a method, named BRAU-Net, to solve the pubic symphysis-fetal head segmentation task. The method adopts a U-Net-like pure Transformer architecture with bi-level routing attention and skip connections, which effectively learns local-global semantic information. The proposed BRAU-Net was evaluated on transperineal Ultrasound images dataset from the pubic symphysis-fetal head segmentation and angle of progression (FH-PS-AOP) challenge. The results demonstrate that the proposed BRAU-Net achieves a comparable final score. The codes will be available at \href{https://github.com/Caipengzhou/BRAU-Net}{https://github.com/Caipengzhou/BRAU-Net}.
\end{abstract}



\begin{keyword}
FH-PS-AOP \sep U-Net \sep Bi-level Routing Attention \sep Transformer
\end{keyword}

\end{frontmatter}


\section{Introduction}
\label{sec:sample1}
The extended duration of labor, resulting from a sluggish fetal descent, is related to a heightened risk of maternal and perinatal morbidity \cite{RN1}. However, accurately determining the fetal head (FH) station through monitoring poses a significant clinical challenge in obstetric management\cite{RN2}. Traditionally, clinical estimation of fetal station primarily relies on transvaginal digital examination, which is subjective, often cumbersome, and lacks reliability\cite{RN3,RN4}.

To address the demand for a more objective diagnosis, transperineal Ultrasound (TPU) has emerged as a viable solution. The TPU facilitates the assessment of FH station by measuring the angle of progression (AoP), representing the extension that the FH undergoes during its descent\cite{RN6}. Numerous studies have indicated that AoP serves as an objective, accurate, dependable, and reproducible parameter for evaluating FH descent, aiding clinicians in their daily decision-making for optimal diagnosis\cite{RN7,RN8,RN9,RN10}.
Manual segmentation of the symphysis pubis (SP) and fetal head from intrauterine images is currently considered the most dependable method, but it is exceptionally time-consuming and susceptible to subjectivity and significant inter-observer variability. \cite{RN11,RN12} proposed a deep learning-based framework for segmenting the region of PS-FH and locating the landmark of PS endpoints from ultrasound standard planes. And then AoP was measured from the central axis and the tangent point .
In this paper, we propose a pubic symphysis-fetal head segmentation method based on bi-level routing attention\cite{RN13}. The rest of the paper is organized as follows. In section 2, we present the method that allows to effective segment pubic symphysis-fetal head in transperineal Ultrasound image. In section 3, we describe the performance of our approach on the final stage test set. Finally, section 4 concludes the paper.
\section{Methods}
\subsection{Bi-Level Routing Attention Mechanism(BRA)}
\textbf{Patch partition and Linear projection.} By dividing a 2D input feature map $X \in \mathbb{R}^{H \times W \times C}$ into $S \times S$ non-overlapped regions, each region contains $\frac{{HW}}{{S^2}}$. Subsequently, the query, key, and value tensor are obtained through linear projection:
\[Q = X{W^q}, K = X{W^k}, V = X{W^v}.\]
where $Q, K, V \in {{\mathbb R}^{H \times W \times C}}$ are query, key, value, respectively. The ${W^q}, {W^k}\\, {W^v} \in{{\mathbb R}^{C \times C}}$
are linear projection weights matrix for the query, key, value, respectively.
\\
\textbf{Region-to-region routing.} The process begins by calculating the average of Q and K for each region respectively, resulting in region-level queries and keys, ${Q^r},{K^r} \in {{\mathbb R}^{{S^2} \times C}}$. Next, the region-to-region adjacency matrix is derived, ${A^r} \in {{\mathbb R}^{{S^2} \times \;{S^2}}}$ via applying matrix multiplication between ${Q^r}$ and transposed ${K^r}$. Entries measure how much two regions are semantically related in the ${A^r}$. The key step involves retaining only the top-k connections for each region. The region-to-region routing can be formulated as:
\[{A^r} = {Q^r}{({K^r})^T}.\]
\[{I^r} = topkIndex({A^r}).\]
Here, ${I^r} \in {{\mathbb N}^{{S^2} \times k}}$ the k indices of most relevant regions for the ${i^{th}}$ region are contained in the ${i^{th}}$
row of ${I^r}$.
\begin{figure}
\centering
\includegraphics[width=5in, keepaspectratio]{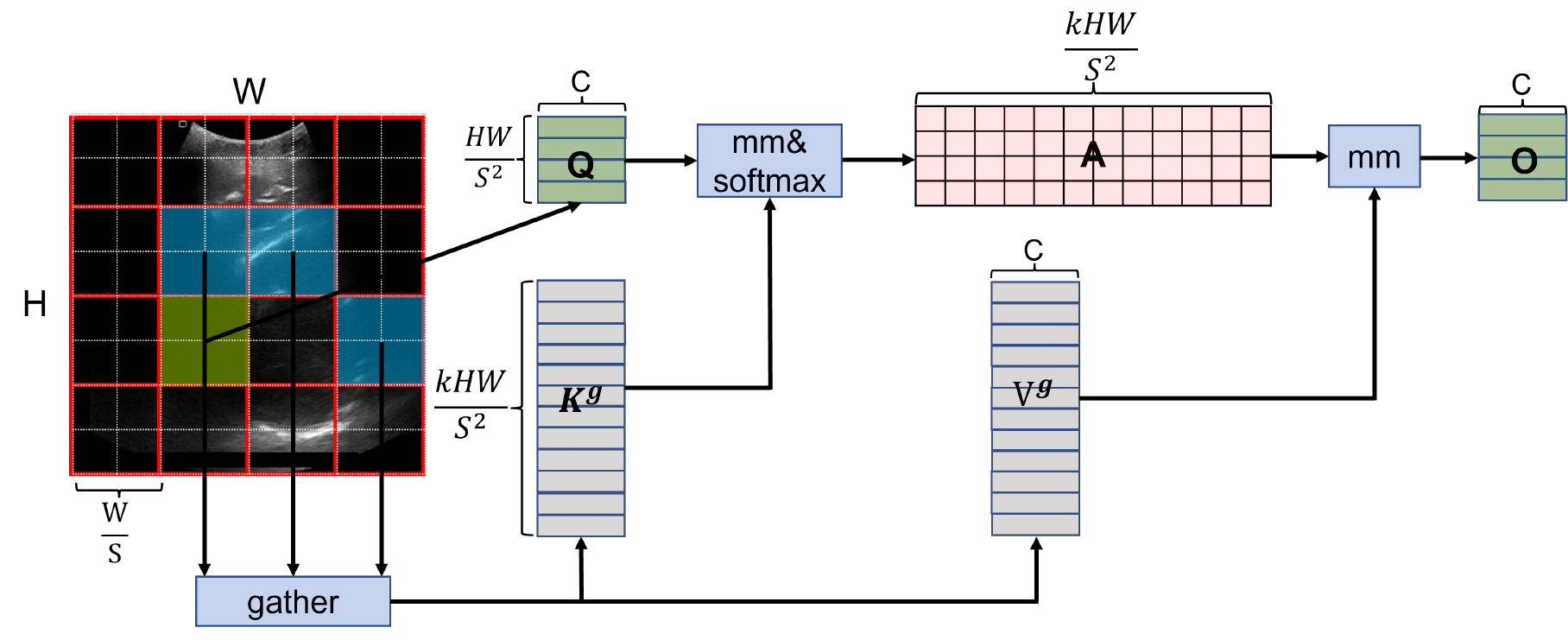}\\
\caption{The token-to-token attention.}
\hypertarget{first_image}{}
\end{figure}
\\
\textbf{Token-to-token attention.} In this section, the first step is to gather key and value tensor. For each query in region, it will attend to all key-value pairs residing in the union of k routed regions. The token-to-token attention is shown in \hyperlink{first_image}{Figure 1}.

\[{K^g} = gather(K,{I^r}), {V^g} = gather(V,{I^r}).\]

\[O = softmax(\frac{{Q{{({K^g})}^T}}}{{\sqrt C }}){V^g} + LCE(V).\]

Where ${K^g},{V^g} \in {{\mathbb R}^{kHW \times C}}$ are gathering key and value tensor. The function LCE(·) is parameterized using a depth-wise convolution with a kernel size set to 5.

\subsection{Architecture overview}
The overall architecture of BRAU-Net is shown in \hyperlink{fig:second_image}{Figure 2(a)}. Inspired by U-Net\cite{RN14}, we design a symmetric transformer-based decoder. The BRAU-Net includes encoder, decoder, bottleneck and skip connections.
The patch embedding module is used for patch partition and linear embedding, which consists of two $3 \times 3$ convolution layer that the first convolution layer is followed by a GELU activation and BatchNorm and another is only followed by BatchNorm. Then, The block can effectively learn local-global semantic information. When down-sampling leads to a decrease in the resolution of the feature map to $\frac{{16}}{H} \times \frac{{16}}{W}$, the number of blocks is 8, while in other stages, the number of blocks is set to 4. Follow as Swin-Unet\cite{RN15}, we use patch merging to decrease resolution of feature map and increase dimension. In the bottleneck, the resolution of the feature map remains unchanged. The extracted context features are fused with multiscale features from encoder via skip connections to complement the loss of spatial information caused by down-sampling. For the decoder, patch expanding is used to restore the resolution of the feature, and $4 \times $ patch expanding is used when the resolution is finally restored to $H \times W$. Subsequently, a linear projection layer is employed to generate pixel-level segmentation predictions from these up-sampled features.

\begin{figure}[H]
\centering
\includegraphics[width=4.5in, keepaspectratio]{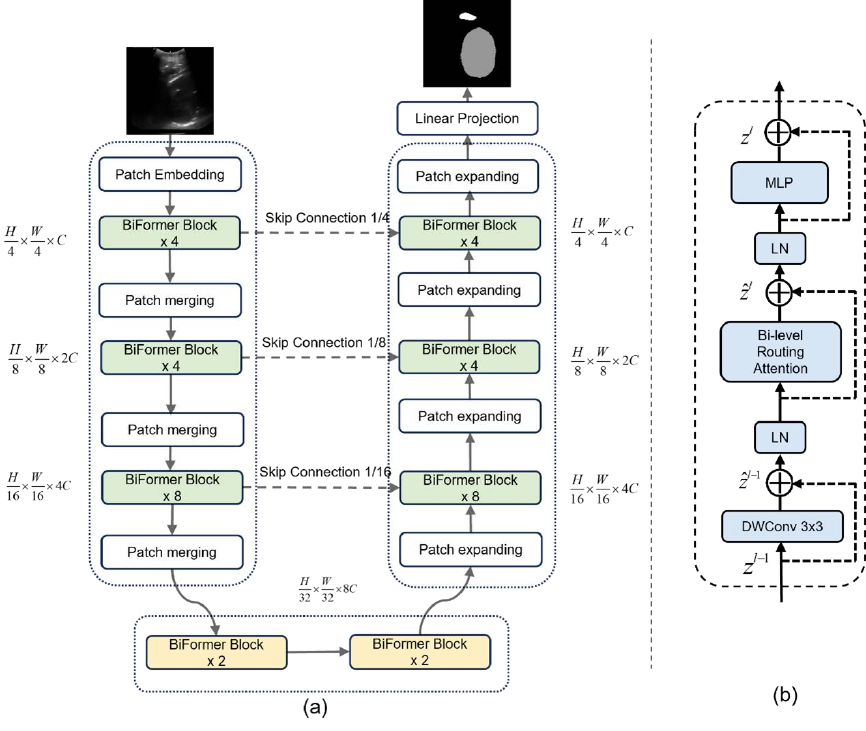}\\
\caption{(a): The architecture of our BRAU-Net, which is constructed based on BiFormer block. (b): Details of a BiFormer Block.}
\hypertarget{fig:second_image}{}
\end{figure}

\subsection{BiFormer Block}
The BRA filters out irrelevant key-value pairs at the coarse-grained area level, and applies fine-grained attention to the federated routing area. The BiFormer block is presented in \hyperlink{fig:second_image}{Figure 2(b)}. The block is built based on BRA, which is convolved by $3 \times 3$ depth-wise convolution, LayerNorm(LN) layer, BRA module, residual connection and 2-layer MLP with expansion ratio e. The $3 \times 3$ depth-wise convolution can implicitly encodes relative position information. The BiFormer block can be formulated as:
\[{\hat z^{l - 1}} = DW({z^{l - 1}}) + {z^{l - 1}}\]
\[{\hat z^l} = BRA(LN({\hat z^{l - 1}})) + {\hat z^{l - 1}}\]
\[{z^l} = MLP(LN({\hat z^l})) + {\hat z^l}\]
where ${\hat z^{l - 1}}$, ${\hat z^l}$ and ${z^l}$ represent the outputs of the Depth-wise convolution, Bi-level Routing Attention module and MLP module of the 
${l^{th}}$ block, respectively.

\section{Experiment and Results}
The BRAU-Net is implemented based on Python 3.7 and PyTorch 1.13.1. All our experiments were conducted on a single NVIDIA GeForce RTX 3060Ti with 8GB. We employed the Adam optimizer optimize our model during the back propagation. The data augmentations such as flips, rotations are used to enhance the diversity of the data. The number of attention heads and topks in each stage are 2, 4, 8, 16 and 4, 8, 16, -2, respectively. We set C to 96 and e to 3, respectively. The BRAU-Net is trained from scratch. Since the test set was not previously published, we randomly split the 4000 cases into separate training(3600 cases) and validation(400 cases) datasets. And the average Dice-Similarity coefficient (DSC) and average Hausdorff Distance(HD) are used as evaluation metric to evaluate our method. The quantitative results for the validation dataset(400cases) are presented in Table 1. We also demonstrate visualization results in \hyperlink{fig:third_image}{Figure 3}.

\begin{table}[htbp]
\centering
\caption{The quantitative results on the validation dataset}
\label{tab:your_table_label}
\begin{tabular}{lllll}
\toprule
Method   & DSC   & HD   & PS    & FH    \\
\midrule
BRAU-Net & 96.59 & 2.19 & 95.42 & 97.76 \\     
\bottomrule
\end{tabular}
\end{table}

Tabel 2 shows the quantitative results on the final testing phase.
The final score is computed as : S=0.25(DSC\_FH+DSC\_PS+DSC\_ALL)/3.0+0.25[0.5(\\1-HD\_FH/100+1-HD\_PS/100+1-HD\_ALL/100)/3.0+0.5(1-ASD\_FH/100+1-ASD\_PS/100+1-ASD\_ALL/100)/3.0]+0.5(1-$\triangle$AOP/180).
Here, FH, PS and ALL, respectively, denote the region of fetal head, Pubic Symphysis and the whole image. DSC, HD, ASD and AOP, respectively, represent Dice-Similarity coefficient, Hausdorff Distance, Average Surface Distance and Angle of Progression. 

\begin{table}[htbp]
\centering
\caption{The quantitative results on the final testing phase.}
\label{tab:my_table}
\begin{tabular}{lccccccc}
\toprule
Method & Score & AOP & HD\_FH & HD\_PS\\ & ASD\_FH & ASD\_PS & HD\_ALL & ASD\_ALL\\ & DSC\_FH & DSC\_PS & DSC\_ALL\\
\midrule
BRAU-Net & 89.74 & 12.20 & 20.03 & 14.07\\ & 7.10 & 4.21 & 21.87 & 6.06\\ & 0.88 & 0.80 & 0.87\\
\bottomrule
\end{tabular}
\end{table}
\begin{figure}
\centering
\includegraphics[width=3in, keepaspectratio]{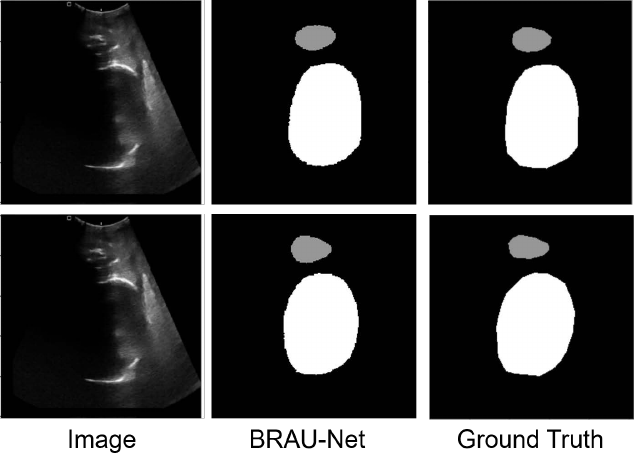}\\
\caption{The some of visualization results on the validation set.}
\hypertarget{fig:third_image}{}
\end{figure}

\section{Conclusion}
In this paper, we introduced a novel pure Transformer-based U-shaped encoder-decoder for pubic symphysis-fetal head segmentation from transperineal Ultrasound images. In order to leverage the power of Transformer, we take BiFormer block as the basic unit for feature representation and long-range semantic information interactive learning. The results demonstrate that the proposed BRAU-Net achieves a preferable final score.

\bibliographystyle{elsarticle-num} 

\begin{thebibliography}{10}
\expandafter\ifx\csname url\endcsname\relax
  \def\url#1{\texttt{#1}}\fi
\expandafter\ifx\csname urlprefix\endcsname\relax\def\urlprefix{URL }\fi
\expandafter\ifx\csname href\endcsname\relax
  \def\href#1#2{#2} \def\path#1{#1}\fi
  
\bibitem{RN1}
M.~Fitzpatrick, K.~McQuillan, C.~J.~O. O’Herlihy, Gynecology, Influence of persistent occiput posterior position on delivery outcome 98~(6) (2001) 1027--1031.

\bibitem{RN2}
P.~J.~B. Simkin, The fetal occiput posterior position: state of the science and a new perspective 37~(1) (2010) 61--71.

\bibitem{RN3}
S.~DM, B.~KS, L.~J. U. i.~O. O, G.~T. O. J. o. t. I. S. o. U.~i. Obstetrics, Gynecology, Intrapartum fetal head position i: comparison between transvaginal digital examination and transabdominal ultrasound assessment during the active stage of labor 19~(3) (2002) 258--263.

\bibitem{RN4}
O.~Dupuis, S.~Ruimark, D.~Corinne, T.~Simone, D.~André, R.~J. E. J. o.~O. René-Charles, Gynecology, R.~Biology, Fetal head position during the second stage of labor: comparison of digital vaginal examination and transabdominal ultrasonographic examination 123~(2) (2005) 193--197.


\bibitem{RN6}
A.~Barbera, X.~Pombar, G.~Perugino, D.~Lezotte, J.~J. U. i.~O. Hobbins, Gynecology, A new method to assess fetal head descent in labor with transperineal ultrasound 33~(3) (2009) 313--319.

\bibitem{RN7}
A.~Dall’Asta, L.~Angeli, B.~Masturzo, N.~Volpe, G.~B.~L. Schera, E.~Di~Pasquo, F.~Girlando, R.~Attini, G.~Menato, T.~J. A. j. o.~o. Frusca, gynecology, Prediction of spontaneous vaginal delivery in nulliparous women with a prolonged second stage of labor: the value of intrapartum ultrasound 221~(6) (2019) 642. e1--642. e13.

\bibitem{RN8}
E.~Montaguti, N.~Rizzo, G.~Pilu, A.~J. T. J. o. M.-F. Youssef, N.~Medicine, Automated 3d ultrasound measurement of the angle of progression in labor 31~(2) (2018) 141--149.

\bibitem{RN9}
E.~Brunelli, A.~Youssef, E.~M. Soliman, B.~Del~Prete, M.~H. Mahmoud, M.~Fikry, G.~Pilu, R.~A. J. A. J. o.~O. Kamel, Gynecology, The role of the angle of progression in the prediction of the outcome of occiput posterior position in the second stage of labor 225~(1) (2021) 81. e1--81. e9.

\bibitem{RN10}
M.~Zhou, C.~Yuan, Z.~Chen, C.~Wang, Y.~Lu, Automatic angle of progress measurement of intrapartum transperineal ultrasound image with deep learning, in: Medical Image Computing and Computer Assisted Intervention–MICCAI 2020: 23rd International Conference, Lima, Peru, October 4–8, 2020, Proceedings, Part VI 23, Springer, pp. 406--414.

\bibitem{RN11}
A.~Youssef, E.~Brunelli, C.~Azzarone, G.~Di~Donna, P.~Casadio, G.~J. U. i.~O. Pilu, Gynecology, Fetal head progression and regression on maternal pushing at term and labor outcome 58~(1) (2021) 105--110.

\bibitem{RN12}
Y.~Lu, D.~Zhi, M.~Zhou, F.~Lai, G.~Chen, Z.~Ou, R.~Zeng, S.~Long, R.~Qiu, M.~J.~C. Zhou, M.~M.~i. Medicine, Multitask deep neural network for the fully automatic measurement of the angle of progression 2022 (2022).

\bibitem{RN13}
L.~Zhu, X.~Wang, Z.~Ke, W.~Zhang, R.~W. Lau, Biformer: Vision transformer with bi-level routing attention, in: Proceedings of the IEEE/CVF Conference on Computer Vision and Pattern Recognition, pp. 10323--10333.

\bibitem{RN14}
O.~Ronneberger, P.~Fischer, T.~Brox, U-net: Convolutional networks for biomedical image segmentation, in: Medical Image Computing and Computer-Assisted Intervention–MICCAI 2015: 18th International Conference, Munich, Germany, October 5-9, 2015, Proceedings, Part III 18, Springer, pp. 234--241.

\bibitem{RN15}
H.~Cao, Y.~Wang, J.~Chen, D.~Jiang, X.~Zhang, Q.~Tian, M.~Wang, Swin-unet: Unet-like pure transformer for medical image segmentation, in: European conference on computer vision, Springer, pp. 205--218.




\end{thebibliography}

\end{document}